\documentclass[%
 reprint,
superscriptaddress,
prb,
floatfix,
]{revtex4-2}

\usepackage{graphicx}% Include figure files
\usepackage{dcolumn}% Align table columns on decimal point
\usepackage{bm}% bold math
\usepackage{natbib}
\usepackage{ulem}
\usepackage{xcolor}
\usepackage{amsmath} %for multyline formulae
\usepackage{amssymb}
\usepackage{hyperref}% add hypertext capabilities
\usepackage{multirow}
\usepackage{enumitem}

\definecolor{newtext}{RGB}{0, 0, 0}

\begin{document}

\preprint{APS/123-QED}

\title{
Switching picosecond magnetoacoustic regimes in a ferromagnetic waveguide
}

\author{P. I. Gerevenkov}
\email{petr.gerevenkov@mail.ioffe.ru}
\homepage{http://www.ioffe.ru/ferrolab/}
 \affiliation{Ioffe Institute, St. Petersburg 194021, Russia}
\author{Ia. A. Mogunov}
 \affiliation{Ioffe Institute, St. Petersburg 194021, Russia} 
\author{Ia. A. Filatov}
 \affiliation{Ioffe Institute, St. Petersburg 194021, Russia}
 \author{N. S. Gusev}
 \affiliation{Institute for Physics of Microstructures, RAS, Nizhny Novgorod 603087, Russia}
 \author{M. V. Sapozhnikov}
 \affiliation{Institute for Physics of Microstructures, RAS, Nizhny Novgorod 603087, Russia}
 \affiliation{Lobachevsky State University of Nizhny Novgorod, Nizhny Novgorod 603022, Russia}
 \author{N. E. Khokhlov}
 %\affiliation{Ioffe Institute, St. Petersburg 194021, Russia}
 \altaffiliation[Currently at ]{Radboud University, Nijmegen, Netherlands}%Lines break automatically or can be forced with \\
\author{A. M. Kalashnikova}
 \affiliation{Ioffe Institute, St. Petersburg 194021, Russia}

\date{\today}% It is always \today, today,

\begin{abstract}
The development of magnonics requires energy-efficient methods for generating spin waves and controlling their parameters. 
Acoustic waves are known to resonantly excite spin waves through magneto-elastic wave formation or to induce non-resonant forced magnetization oscillations. 
Short acoustic wavepackets enable another unexplored resonant interaction regime -- the Cherenkov radiation of spin waves. 
This raises a question regarding the criteria and signatures of these three regimes of picosecond magnetoacoustics \textcolor{newtext}{and transitions between them in confined magnetic structures}.
Here, we use a scanning magneto-optical pump-probe technique to directly observe all three regimes of interaction between laser-driven acoustic and magnetostatic wavepackets in a thin permalloy film and a waveguide fabricated on Si substrate. 
Direct measurements of the phase velocities reveal the transition from a coupled magneto-elastic wavepacket to Cherenkov-like radiation and the non-resonant regime, controlled by the detuning between magnon and phonon group velocities.
The acoustic pulse is found to be affected by the excited magnetization dynamics only in the magneto-elastic regime.
\end{abstract}

\maketitle

\textcolor{newtext}{\section{\label{sec:intro} Introduction}}

Increasing demands for computing efficiency drive the search for new ways of processing information.
In recent years, prototypes and concepts for constructing full-fledged magnonic computing devices have emerged~\cite{mahmoud2021spin,garlando2023numerical}.
In particular, it has been demonstrated~\cite{mahmoud2020introduction} that available magnonic elements such as half-adder~\cite{wang2020magnonic}, diode~\cite{szulc2020spin} and majority gate~\cite{talmelli2020reconfigurable} form a sufficient set for conventional digital computing.
Several unconventional computing methods employing magnonics are also being actively developed, including neuromorphic and reservoir computing~\cite{papp2021nanoscale,watt2021enhancing,balynsky2021quantum,korber2023pattern}.
All these approaches rely heavily on magnetostatic spin waves.
Recent micromagnetic simulations predict that magnonic computation can outperform classical 7-nm CMOS technologies~\cite{mahmoud2022would}, but this requires improved spin-wave excitation efficiency, pulsed operation, and seamless integration with existing CMOS and electronic components.

One of the promising directions to meet these requirements is to utilize the coupling between spin waves and a surface acoustic waves (SAW).
\textcolor{newtext}{The studies of magneto-elastic interaction underlying this coupling have a long history, with pioneering works demonstrating mutual excitation of spin and acoustic waves in ferromagnetic films~\cite{pomerantz1961excitation, kooi1963interaction, wigen1965coherent}.}
This interaction enables acoustic control of spin wave parameters with superior energy efficiency (see ch.~8 of \cite{barman20212021}, ch.~10 of \cite{delsing20192019}, and \cite{Vlasov2022_magnetoac_review,Yang2021_review,Li2021mag-saw-review,bukharaev2018straintronics}) and selectivity to various acoustic modes~\cite{Yamamoto2022_SW-SAW_interaction_theor,babu2020interaction,Hwang2024couplMagAc}.
Such a coupling is now actively explored in the modern field of picosecond magnetoacoustics~\cite{scherbakov2023ultrafast}, where short acoustic pulses enable novel dynamics.
When short acoustic pulses are used, the spin wave-SAW coupling becomes a source of nonlinearity that does not require large dynamic amplitudes and can be harnessed for reservoir computing~\cite{yaremkevich2023chip}.

The propagation of an acoustic pulse in a magnetic layer can lead to two distinct resonant effects. 
The first is the formation of a coupled magneto-elastic wave (MEW), also called a magnon-polaron. 
This effect has been demonstrated both in bulk systems~\cite{Bombeck2012acSW_in_GaMnAs,Hashimoto2018_coupl_in_garnet} and at surfaces~\cite{Casals2020magac_wave,Kuss2020non-repr_magac_wave,hioki2022coherent,Hwang2024couplMagAc}. 
The second is spin wave excitation through a Cherenkov-like process. 
This effect was predicted in micromagnetic simulations for a general localized moving field pulse~\cite{yan2013spin}, and was only recently demonstrated in picosecond acoustics for a bulk magnet~\cite{Filatov2024Tobe}.
In both effects, the energy transfer between spin wave and lattice oscillations is realized through the same magnetoelastic coupling.
The aforementioned fact raises a fundamental question about the distinction between these two interation regimes.
\textcolor{newtext}{In addition to these two resonant regimes, a non‑resonant regime of forced magnetization oscillations can also occur when the acoustic and spin‑wave dispersions do not intersect.}

\textcolor{newtext}{In this Article, using laser-induced short acoustic wavepackets, we experimentally demonstrate all three regimes of magnetoacoustic interaction in a single system: Cherenkov-like spin-wave emission, magnetoelastic wavepacket formation, and non-resonant forced magnetization oscillations.
By means of a theoretical model, we identify the conditions for observing each regime and the transitions between them in terms of the group-velocity detuning between the acoustic and spin-wave components.
Furthermore, employing a magnonic structure with a nonuniform magnetization direction (a T-shaped waveguide), we experimentally demonstrate local switching between the regimes during the propagation of a single acoustic pulse.}

\textcolor{newtext}{The article is organized as follows. 
Section~\ref{sec:experimental} describes the studied structures, a continuous Permalloy (Py) film and a T‑shaped Py waveguide on a single‑crystal Si substrate, and the femtosecond scanning optical pump-probe technique. 
Section~\ref{sec:film} presents the experimental observation of the three picosecond magnetoacoustic regimes in the continuous film. 
Section~\ref{sec:theory} provides the theoretical analysis that links the transition between the regimes to the group‑velocity mismatch between the spin and acoustic waves. 
Using the experimental and theoretical results from the preceding Sections, Section~\ref{sec:T-structure} demonstrates the switching between the regimes during the propagation of an acoustic pulse in a T‑shaped waveguide with a nonuniform magnetization direction. 
The Conclusion summarizes the main findings and outlines their potential relevance for energy‑efficient magnonic computing.}

\textcolor{newtext}{\section{\label{sec:experimental} Methods}}

\textcolor{newtext}{
Two kinds of samples were used in the study.
A continuous film with uniform magnetization is used to investigate various regimes of picosecond magnetoacoustics, while a T-shape waveguide with local magnetization nonuniformities is used to achieve a switching between them.
A polycrystalline permalloy (Ni$_{80}$Fe$_{20}$, Py) film of thickness $d = 20$\,nm is deposited on a monocrystalline (011)-Si substrate by magnetron sputtering (Sec. I in Suppl. Materials~\cite{Supplemental}).
Py serves as a model magnonic material with a low Gilbert damping $\alpha_G = 0.01$~\cite{zhao2016experimental, han2009magnetic}, allowing clear observation of propagating spin waves.
Its saturation magnetization $M_S = 8 \cdot 10^5$\,A\,m$^{-1}$~(\cite{khokhlov2021neel,vansteenkiste2014design} and Sec.~I in Suppl. Materials~\cite{Supplemental}) is lower than that of typical magnetoacoustic materials such as CoFeB~\cite{shelukhin2020laser} or FeGa~\cite{atulasimha2011review}, which results in lower spin-wave frequencies and thereby brings the intersection point of the acoustic and spin-wave dispersions into the experimentally accessible wavenumber range.
The polycrystalline nature of Py ensures no in-plane magnetocrystalline anisotropy, so the equilibrium magnetization $\mathbf{M}$ is collinear with the external in-plane magnetic field $\mathbf{H}_{ext}$ ($\mathbf{M} \parallel \mathbf{H}_{ext}$) in the continuous film.
In the T-shape waveguide with a nonuniform magnetization direction, the local $\mathbf{M}$ is determined by the sum of $\mathbf{H}_{ext}$ and the shape anisotropy field, and remains strictly in-plane.
Despite the relatively small magnetoelastic coefficients of Py~\cite{adhikari2026optical}, these coefficients depend strongly on the composition~\cite{pomerantz1961excitation} and are non-zero without special sample preparation.
The conditions described above enable the observation of magnetoacoustic interaction in the Py/Si system, as demonstrated, for example, in Refs.~\cite{pomerantz1961excitation,adhikari2026optical} and in the present work.}

\textcolor{newtext}{
From the Py film we fabricate structures with high edge quality using maskless laser writing: a square of lateral size 500\,$\mu$m with a homogeneous magnetization distribution, and a T‑shaped waveguide of width 5\,$\mu$m~(Sec.~I in Suppl. Materials~\cite{Supplemental}), in which a 90\,$^\circ$ rotation of the magnetization occurs over a certain range of applied fields due to the spatially nonuniform shape anisotropy.
The dispersion of acoustic waves in the studied frequency range is taken as linear, which agrees with numerical simulations.
The spin-wave dispersion in the continuous film and in the T-shaped waveguide is obtained from micromagnetic simulations using the mumax$^3$ package~\cite{vansteenkiste2014design} (Sec.~II in Suppl. Materials~\cite{Supplemental}).}

\begin{figure}
	\includegraphics[width= \linewidth]{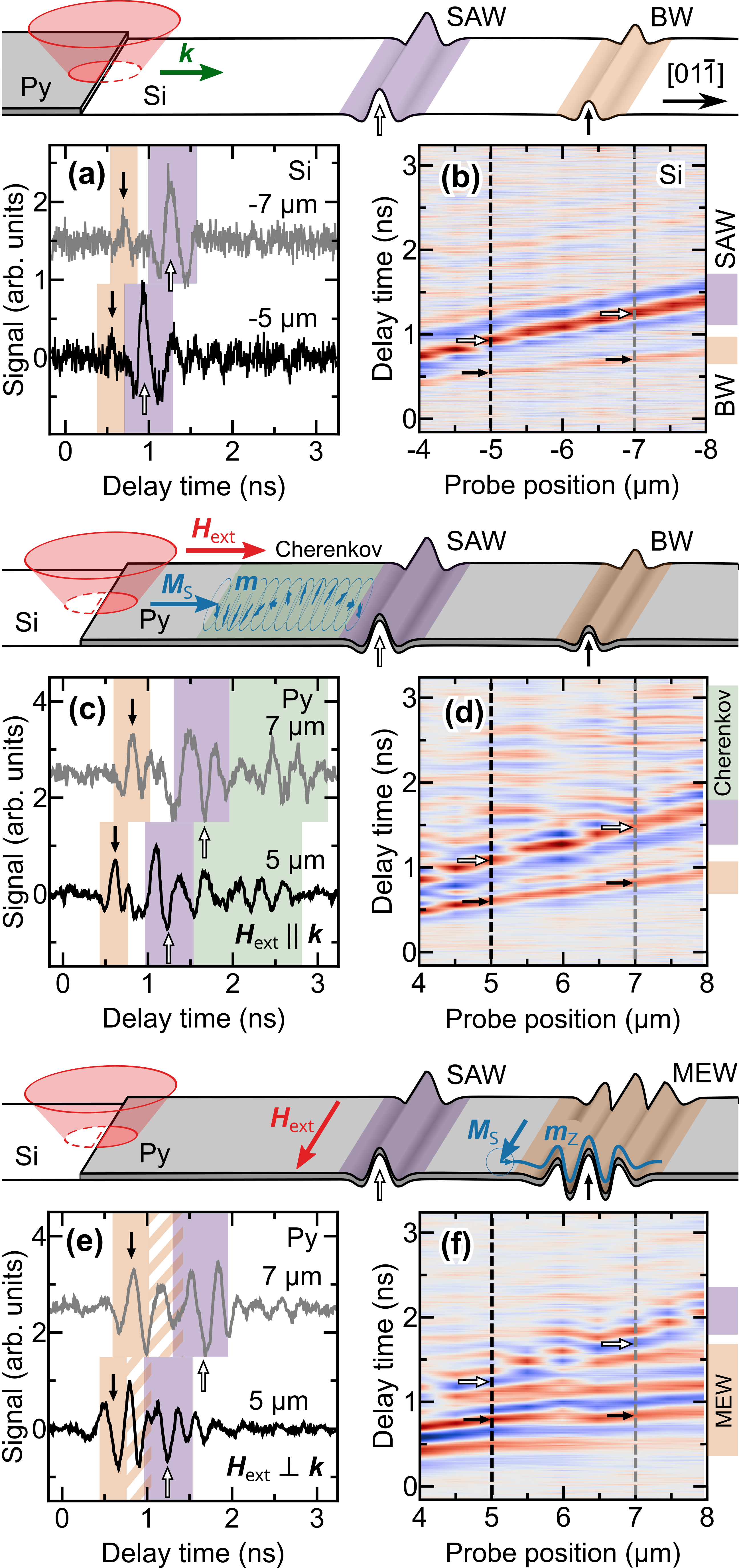}
	\caption{\label{fig:SAW_MSW_Si_NiFe}
		{\bf Acoustic and magnetostatic waves in the Py film on the Si substrate.}
		(a,b) Acoustic wavepackets in the bare Si substrate excited at the Py edge (see sketch at the top of the figure):
		(a) pump-probe polarization rotation signals at different distances from the pump area;
		(b) spatio-temporal map.
		Colors and arrows indicate wavepackets with different group velocities and their centers.
		(c--f) Same for the Py film with $\mathbf{H}_{ext} = 5$\,mT.
		(c,d) Field parallel to the propagation direction (BVSW geometry, Cherenkov-like regime, see sketch above).
		The green area in (c) highlights Cherenkov emission of backward volume spin waves.
		(e,f) Field perpendicular to the propagation direction (SSW geometry, magneto-elastic wavepacket regime, see sketch above).
		The broadening of the faster wavepacket [striped area in (e)] indicates magnetoelastic wavepacket formation.
	}
\end{figure}

\textcolor{newtext}{
For excitation and detection of propagating spin and acoustic waves, we employ a two-colour optical pump–probe technique with temporal and spatial resolution.
Tightly focused pump (wavelength 680\,nm, FWHM = 0.9\,$\mu$m) and probe (wavelength 525\,nm, FWHM = 0.5\,$\mu$m) pulses of 200\,fs duration are incident normally.
This enables detection of waves with wavenumbers up to $k = 1.1\,\mu$m$^{-1}$~\cite{khokhlov2019optical, jackl2017magnon}.
For temporal and spatial resolution, the probe pulses were delayed by a variable delay time and scanned laterally with respect to the pump ones (see Sec.~III in Suppl. Materials~\cite{Supplemental}).
Acoustic waves are generated by thermoelastic expansion of the Py film: no acoustic signal is detected when exciting the bare Si substrate under identical conditions.
The pump spot is placed at the edge of the Py structure (Fig.~\ref{fig:SAW_MSW_Si_NiFe}), which extends a range of wavenumbers of the generated SAW's beyong the one limited by the pump spot diameter. 
The spectra of generated wavepackets are centered at 3.3\,GHz.
For three acoustic wave modes discussed in the next Section, the corresponding central wavenumbers are: 0.3\,$\mu$m$^{-1}$ (bulk wave, BW), 0.6\,$\mu$m$^{-1}$ (Love surface acoustic wave, L-SAW), and 0.8\,$\mu$m$^{-1}$ (Rayleigh surface acoustic wave, R-SAW) (Sec.~III in Suppl. Materials~\cite{Supplemental}).
}

\textcolor{newtext}{\section{\label{sec:results_discussion} Results and Discussion}}

\textcolor{newtext}{\subsection{\label{sec:film} Picosecond Magnetoacoustics in a continuous film}}

To determine the generated SAW modes, we detect wavepackets propagating from the pump spot in a bare (011)-Si substrate.
The signals measured at two distances from the excitation area are shown in Fig.~\ref{fig:SAW_MSW_Si_NiFe}\,(a).
Two short wavepackets propagate with different group velocities (highlighted by colored areas).
The wavepackets are observed in both rotation and reflectivity data (Sec.~IV Suppl. Materials~\cite{Supplemental}) and their shapes do not change during propagation [Fig.~\ref{fig:SAW_MSW_Si_NiFe}\,(b)], indicating the absence of group velocity dispersion~\cite{filatov2022spectrum}. 
This allows us to directly measure the phase velocity by tracking a fixed point in the waveform [indicated by arrows in Fig.~\ref{fig:SAW_MSW_Si_NiFe}\,(a,b)].
For bare Si substrate in the $[01\bar{1}]$ direction, one expects a Rayleigh mode (R-SAW) with elliptically polarized atomic motion in a plane parallel to the acoustic wave vector $k_{\text{SAW}}$ and the surface normal~\cite{book_SAW,Tarasenko2021} (Sec.~V in Suppl. Materials~\cite{Supplemental}).
It is also common to detect a bulk longitudinal acoustic wave (BW) propagating under the surface, with atomic motion similar to that of the R-SAW~\cite{SaitoOptSAWDetect,Sugawara_opt_SAW}.
Accordingly, we classify the packet that propagates at a lower velocity (6.0~km s$^{-1}$) as a R-SAW and the one that propagates at a higher velocity (10.3~km s$^{-1}$) as a BW.
\textcolor{newtext}{The weak oscillations following the R-SAW packet in Fig.~\ref{fig:SAW_MSW_Si_NiFe}\,(b) have the same frequency content as the R-SAW and are attributed to acoustic reflections (Sec.~IV in Suppl. Materials~\cite{Supplemental}).  
Their amplitude is negligible as compared to the discussed signals and does not affect the analysis.}

Now we consider wave propagation from the same excitation spot in the Si substrate capped with a ferromagnetic film [Fig.~\ref{fig:SAW_MSW_Si_NiFe}\,(c-f)].
The presence of the Py film enables an additional Love acoustic mode (L-SAW) with linearly polarized in-plane atomic motion perpendicular to $k_{\text{SAW}}$~\cite{book_SAW} (Sec.~IV, V in Suppl. Materials~\cite{Supplemental}).
\textcolor{newtext}{It also modifies the Rayleigh wave into a quasi‑Rayleigh wave, which behaves similarly to a pure Rayleigh wave when propagating along high-symmetry crystallographic directions of Si.}
The ferromagnetic Py layer also enables surface (SSW) and backwards volume (BVSW) magnetostatic waves~\cite{mahmoud2020introduction}, whose polarization $\mathbf{m}$ depends on the equilibrium magnetization direction $\mathbf{M_{\text{S}}}$~\cite{damon1961magnetostatic,babu2020interaction}.

In the experiment, we again observe two acoustic wavepackets [Fig.~\ref{fig:SAW_MSW_Si_NiFe}\,(c,e)] with velocities slightly different from those in the bare Si substrate due to mechanical loading by the Py layer~\cite{FARNELL1972,Supplemental}.
For $\mathbf{H_{\text{ext}}} || \mathbf{k}$ (BVSW geometry), pronounced oscillations follow the SAW wavepacket [marked by a green area in Fig.~\ref{fig:SAW_MSW_Si_NiFe}\,(c,d)].
The emergence of a BVSW following an acoustic pulse corresponds to the Cherenkov-like emission of a spin wave, recently reported for bulk acoustic wavepackets~\cite{Filatov2024Tobe}.
For $\mathbf{H_{\text{ext}}} \perp \mathbf{k}$ (SSW geometry), such emission is not observed. 
Instead, a significant broadening of the BW packet appears [indicated by the striped area in Fig.~\ref{fig:SAW_MSW_Si_NiFe}\,(e,f)].
\textcolor{newtext}{The broadening may indicate the nonlinear dispersion of the formed magneto-elastic wavepacket (MEW), which causes different frequency components to travel at different phase velocities~\cite{filatov2022spectrum}.}
\textcolor{newtext}{Indeed, the measured phase velocity varies across the extent of the single BW packet, taking values from 20.3 to 39\,km\,s$^{-1}$.}
\textcolor{newtext}{Additionally, the signal-to-noise ratio is significantly higher for all wavepackets compared to the bare substrate case, which we attribute to magneto-optical detection mediated by non-resonant magnetization dynamics in the ferromagnetic layer.}

\textcolor{newtext}{\subsection{\label{sec:theory} Theoretical Model of Interaction Regimes}}

\begin{figure}
	\includegraphics[width=\linewidth]{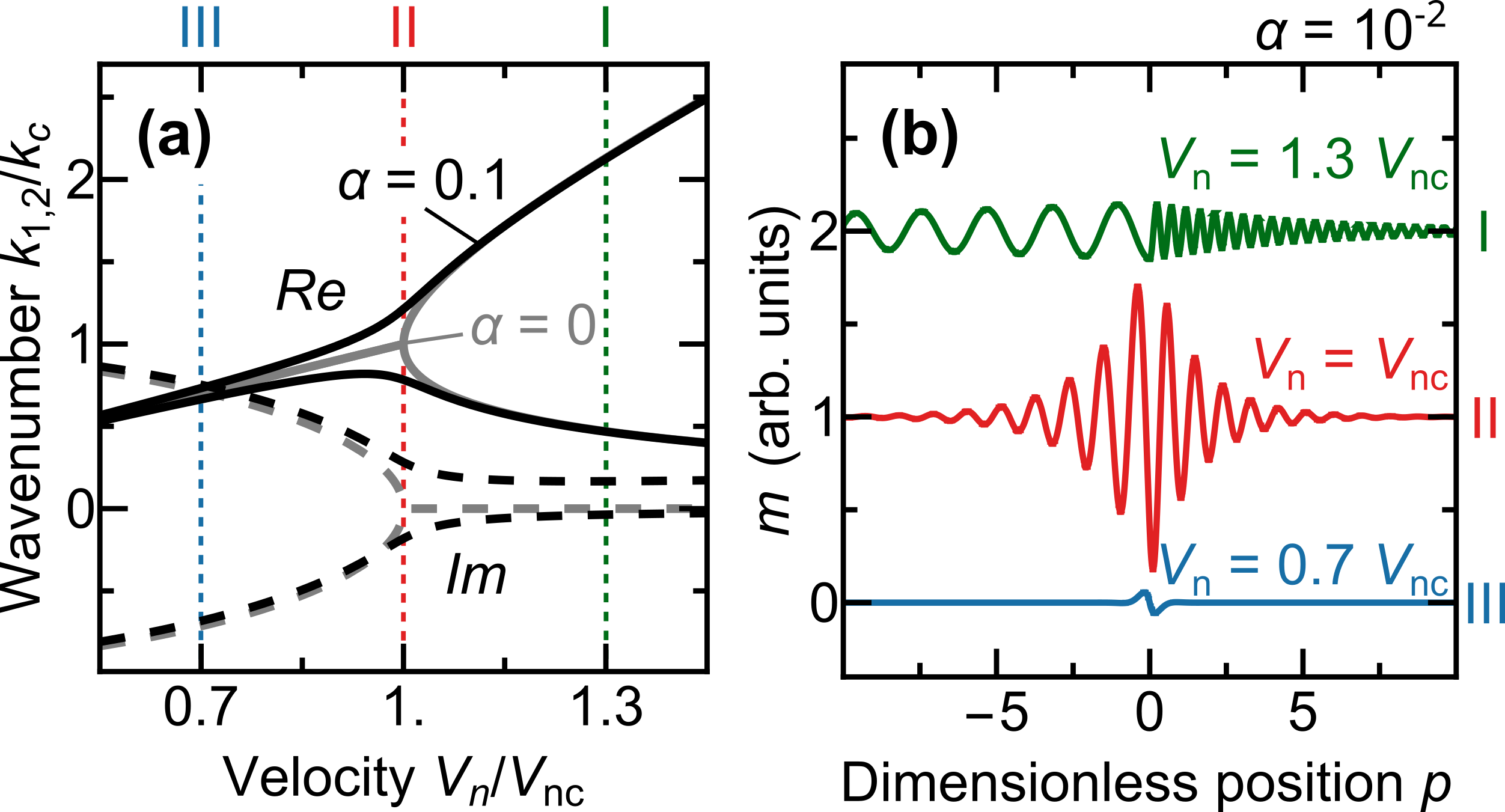}
	\caption{\label{fig:theory}
		\textcolor{newtext}{{\bf Magnetization dynamics vs. group velocity detuning.}
			(a) Real and imaginary parts of the excited wave vectors $k_{1,2}$ of magnetization dynamics from Eq.~(\ref{eq:cherenkov}) vs. acoustic pulse velocity $v_n/v_{nc}$, where $v_{nc}$ is the critical velocity.
			Results are shown for low (gray) and high (black) Gilbert damping.
			(b) Magnetization dynamics $m$ excited by an acoustic pulse modeled as a moving Dirac $\delta$-function, for pulse velocity above (green), equal to (red), or below (blue) the critical value, plotted versus dimensionless distance $p = \tilde{x} k_c$.}
	}
\end{figure}

\begin{figure*}
	\includegraphics[width= 0.9\linewidth]{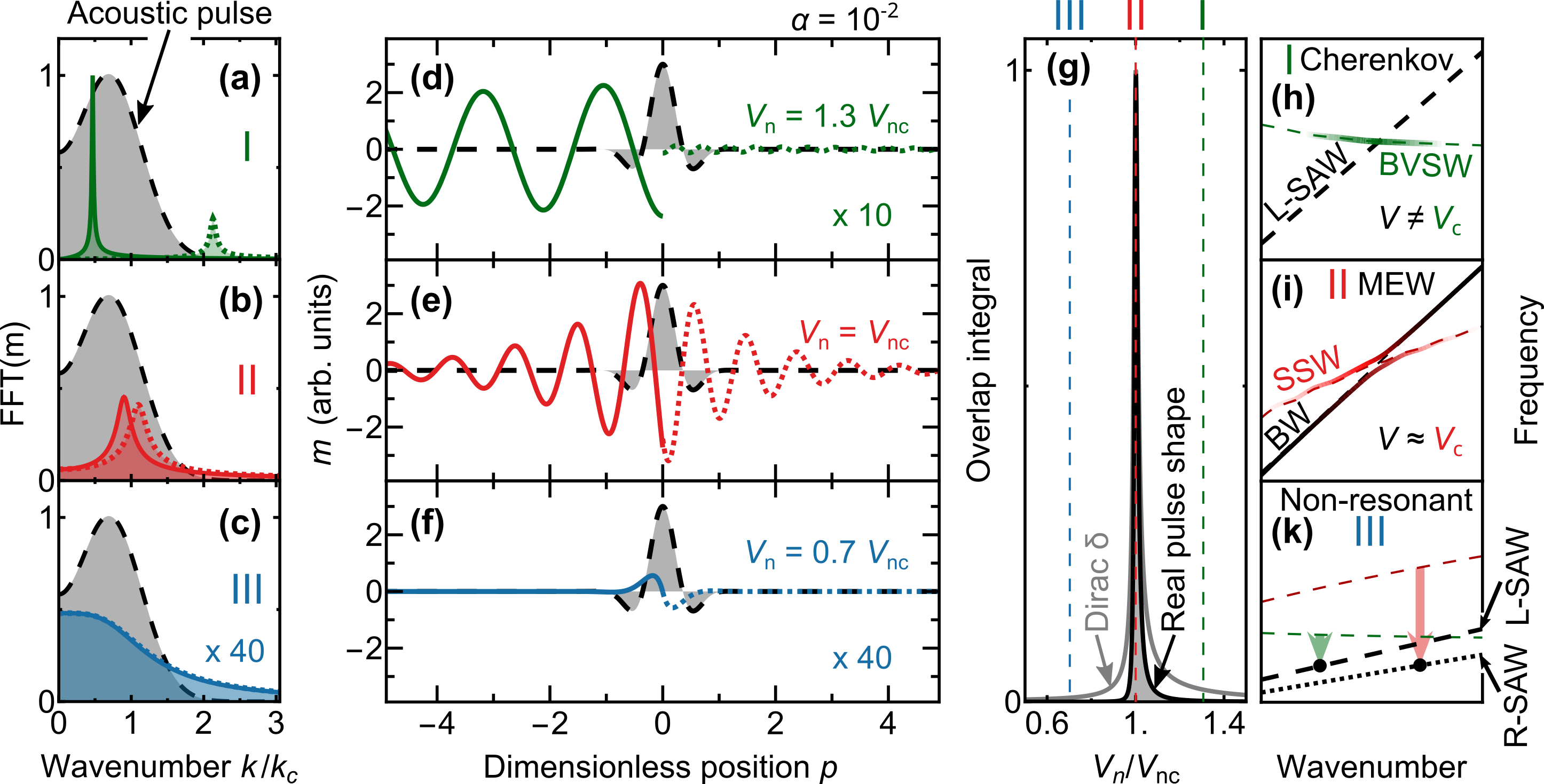}
	\caption{\label{fig:Pulse_shape_and_polarization}
		\textcolor{newtext}{{\bf Influence of the acoustic pulse shape on the excitation regimes.}
			(a–c) Spectra of magnetization dynamics for the three cases from Fig.~\ref{fig:theory}\,(b): $v_n = 1.3\,v_{nc}$ (I, green), $v_n = v_{nc}$ (II, red), and $v_n = 0.7\,v_{nc}$ (III, blue).
			Solid and dashed lines correspond to the two excited wave vectors $k_1$ and $k_2$, respectively.
			The black dashed line shows the spectrum of the acoustic pulse (centered at $k = 0.7 k_c$).
			(d–f) Corresponding spatial profiles $m$ after convolution with the acoustic pulse shape.
			(g) Overlap integral between the acoustic and magnetic amplitudes vs. $v_n/v_{nc}$.
			Vertical dashed lines mark the three cases I (green), II (red), and III (blue) shown in panels (a–f).
			(h–k) Schematic dispersion diagrams illustrating the three regimes, taking into account mode selectivity.
			(h) Cherenkov regime (case I): intersection of the L-SAW and BVSW dispersions.
			(i) Magneto-elastic wavepacket regime (case II): intersection of the BW and SSW dispersions.
			(k) Non-resonant regime (case III): R-SAW and L-SAW dispersons have no crossing with those of BVSW and SSW at accessible wavevectors. Their pairwise non-resonant coupling indicaded by arrows is defined by the selectivity discussed in Sec.~\ref{sec:Selectivety}.}
	}
\end{figure*}

To support the suggested scenarios, we employ a theoretical analysis.
The impact of BW/SAW wavepacket on the magnetization of Py layer is approximated by an effective Landau-Lifshitz (LL) torque $T(x)$ via magnetoelastic coupling (Sec.~VI of Suppl. Material~\cite{Supplemental}).
The magnitude of the LL-torque depends on the relative polarizations of the acoustic and magnetostatic waves~\cite{babu2020interaction}, while their overlap across the film thickness is assumed to be maximal, as $k d \ll 1$~\cite{berk2019strongly}.
\textcolor{newtext}{The torque's spatial shape $T(x)$ along its propagation direction $x$ is given by the acoustic pulse profile.
For the sake of generality, in this part of the analysis we disregard the specifics of the LL-torque for particular acoustic and spin wave modes, assuming it to be nonzero.} 
The torque moves at acoustic velocity $v$ through a medium characterized by a spin-waves dispersions $f(k)$. 
We use the Taylor expansion for $f(k)$ near the point $k_c$ of the tangent intersection of the acoustic and spin wave dispersion. 
The corresponding critical velocity is $f'(k_c) = f(k_c)/k_c = v_c$.
We then obtain the solution for dynamic magnetization $m$:
\begin{align}
\begin{split}\label{eq:cherenkov}
    m(\tilde{x}) = {} &T(\tilde{x}) * \dfrac{1}{k_1 - k_2} \{ e^{2 \pi \imath k_1 \tilde{x}} \Theta \left[\tilde{x} \,\, S_{\Im}(k_1) \right] S_{\Im}(k_1) \\
   & - e^{2 \pi \imath k_2 \tilde{x}} \Theta \left[ \tilde{x} \,\, S_{\Im}(k_2) \right] S_{\Im}(k_2) \}, \\
    k_{1,2} = k_c {} &\left( \Xi (v_n) \pm \sqrt{\Xi (v_n)^2 -1} \right), \\
    \Xi (v_n) = {} &v_n \dfrac{1 + \imath \, \alpha}{1 + \alpha^2} - v_{nc} + 1 ,
\end{split}
\end{align}
where $\tilde{x} = x - v t$ is the distance from the center of the LL-torque profile in the coordinate system moving with the velocity $v$, the asterisk denotes convolution, $k_{1,2}$ are the wavenumbers of two spin-waves, $\Theta ( x )$ is the Heaviside step-function, $S_{\Im}(k)$ denotes the sign of Im($k$), $\alpha$ is the Gilbert damping parameter, and $v_n = v / [ k_c f''(k_c) ]$ and $v_{nc} = v_c / [ k_c f''(k_c) ]$ are dimensionless acoustic velocities.

\textcolor{newtext}{Equation~(\ref{eq:cherenkov}) describes the magnetization dynamics driven by the acoustic pulse without taking into account the back-action on the pulse itself.
We first consider the simplest case of a $T(x)$ in a form of Dirac $\delta$-function pulse.
The real and imaginary parts of the excited wavenumbers as functions of the ratio of the acoustic and spin-wave group velocities are shown in Fig.~\ref{fig:theory}\,(a).
Figure~\ref{fig:theory}\,(b) plots the spatial dynamics of $m$ for three values of this ratio, illustrating the distinct excitation regimes.
In case~I, the dispersions intersect but the group velocities differ significantly.
The excited spin waves are radiated away from the pulse: waves with wave numbers $k_1$ and $k_2$ propagate ahead of or behind the driving pulse according to their group velocities~\cite{yan2013spin,xia2016spin}.
Their spatial decay is governed by the intrinsic Gilbert damping $\alpha$ and the frequency.
In case~II, the condition $v \approx v_{\text{c}}$ holds at the dispersions crossing point, meaning that the pulse and the spin waves have nearly equal group velocities.
This leads to a localization of the excited magnetization dynamics around the acoustic pulse and a resonant enhancement of its amplitude.
In case~III, the dispersions do not intersect and only a non-resonant, low-amplitude magnetization dynamics is observed, which remains confined within the acoustic pulse.}

\begin{figure*}
	\includegraphics[width= 0.95\linewidth]{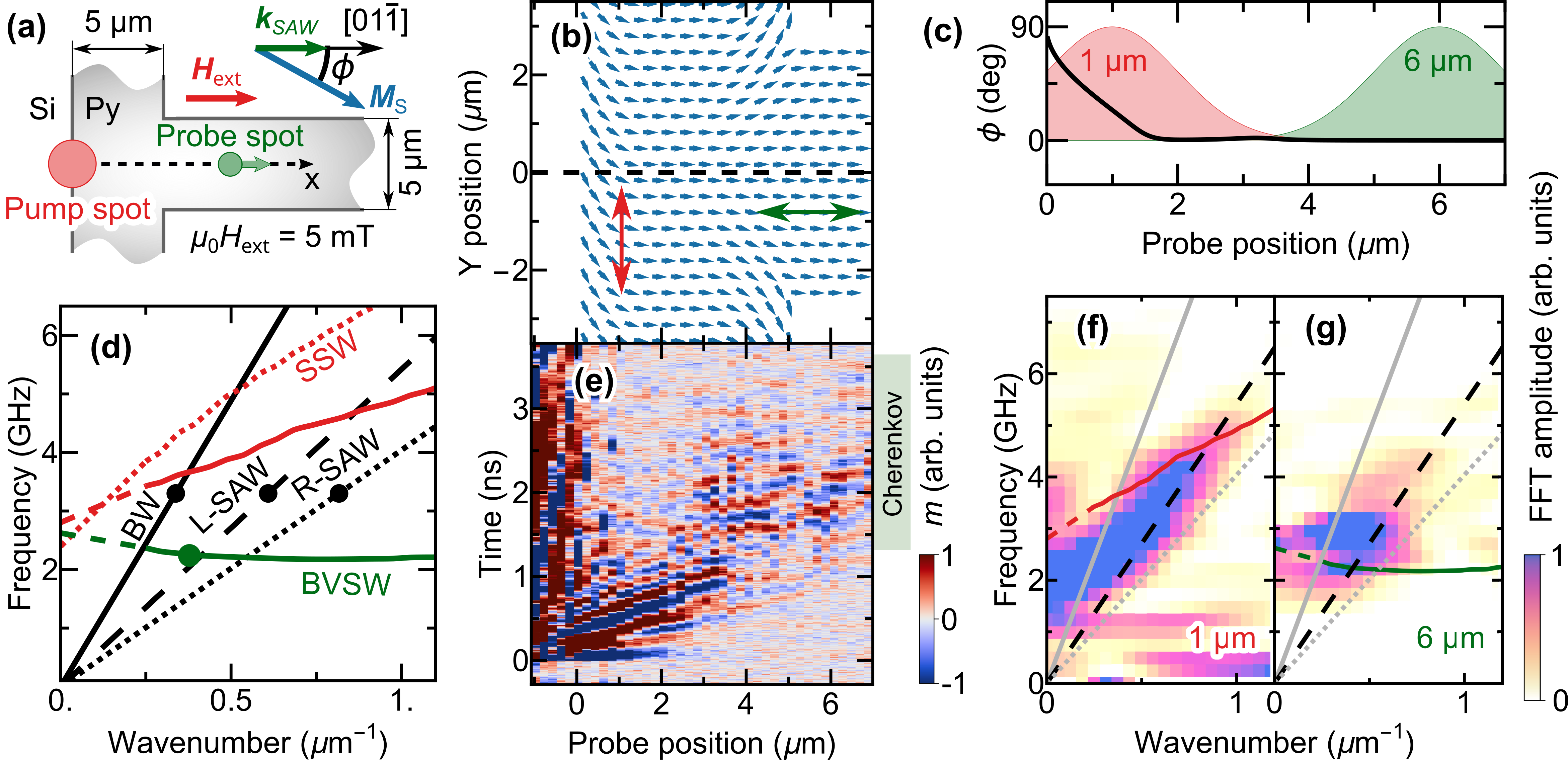}
	\caption{\label{fig:SAW_MSW_T-wg}
		\textcolor{newtext}{{\bf Switching from the non-resonant (III) to the Cherenkov (I) regime in the T-shaped waveguide.}
			(a) Experimental sketch and definition of the angle $\phi$ between $\mathbf{k}$ and local $\mathbf{M}_S$.
			(b) In-plane magnetization distribution in the T-waveguide at $\mu_0 H_{\text{ext}} = 5$\,mT applied along the waveguide axis from micromagnetic simulation.
			The black dashed line indicates the scanning trajectory; double arrows show the shape anisotropy direction in different parts of the waveguide.
			(c) Magnetization deflection angle $\phi$ vs. distance from the waveguide edge along the scanning trajectory [black line in (b)].
			Colored areas show Gaussian spatial windows centered at $x = 1$\,$\mu$m (red) and $x = 6$\,$\mu$m (green) with FWHM of 2.4\,$\mu$m used for local dispersion extraction.
			(d) Dispersion of acoustic and magnetic modes.
			Black lines: BW (solid), L-SAW (dashed), R-SAW (dotted); black dots mark the central wavenumbers of the observed acoustic packets.
			Green line: BVSW dispersion around $x = 6$\,$\mu$m where $\mathbf{M} \parallel \mathbf{H}_{\text{ext}}$; green dot marks the L-SAW -- BVSW intersection where Cherenkov emission occurs.
			Red solid line: spin-wave dispersion around $x = 1$\,$\mu$m, showing an intermediate character between BVSW and SSW (red dotted line).
			(e) Spatio-temporal map of the magneto-optical response. 
			(f,g) Local dispersions obtained from windowed 2D FFT of the data in (e), using the windows shown in (c):
			(f) at $x = 1$\,$\mu$m (non-resonant);
			(g) at $x = 6$\,$\mu$m (Cherenkov).
			Black (gray) lines correspond to the acoustic modes for which coupling to the magnetization is experimentally detected (absent) in the respective geometry; the corresponding magnetostatic dispersions are shown as colored lines.}
	}
\end{figure*}

\textcolor{newtext}{Increasing the Gilbert damping $\alpha$ broadens the transition between the regimes over a wider range of group-velocity mismatch, as seen by comparing the black ($\alpha=0.1$) and gray ($\alpha \approx 0$) solid lines in Fig.~\ref{fig:theory}\,(a).
The effect is most pronounced for regime~II, where increased damping suppresses the peak amplitude of the magnetization dynamics at $v \approx v_{\text{c}}$ [red line in Fig.~\ref{fig:theory}\,(b)].}
\textcolor{newtext}{Thus, in low-damping magnetic insulators such as yttrium iron garnet, one would expect both sharper transitions and a dramatically amplified contrast between the regimes.}
We note that the enhancement of the spin wave amplitude upon tuning the velocity of the localized source to $v \approx v_c$ was reported in~\cite{dobrovolskiy2021cherenkov}.
\textcolor{newtext}{The conditions for the three regimes can be generalized beyond the vicinity of $k_\text{c}$, and applied to the experimental observations.}

\textcolor{newtext}{Accounting for a realistic finite‑size acoustic pulse [black dashed line in Fig.~\ref{fig:Pulse_shape_and_polarization}\,(d-f)] modifies the amplitudes of the excited spin waves according to their spectral overlap with the driving pulse [Fig.~\ref{fig:Pulse_shape_and_polarization}\,(a-c)].
This leads to a reduction of the amplitude in regimes~I and~III [Fig.~\ref{fig:Pulse_shape_and_polarization}\,(d,f)] compared to regime~II [Fig.~\ref{fig:Pulse_shape_and_polarization}\,(e)].
As a result, the magnetization dynamics amplitude inside the acoustic pulse (overlap integral) exhibits a sharp peak at regime~II [Fig.~\ref{fig:Pulse_shape_and_polarization}\,(g)].
For comparison, the gray line in Fig.~\ref{fig:Pulse_shape_and_polarization}\,(g) shows the result for a finite pulse but without the spectral‑overlap correction (equivalent to excitation by a Dirac $\delta$‑function).}

\textcolor{newtext}{We now relate the three identified regimes to the experimental observations in Fig.~\ref{fig:SAW_MSW_Si_NiFe}:}
\begin{enumerate}[label=\Roman*.]
    \item The L-SAW and BVSW dispersion curves intersect, and $v \neq v_\text{c}$ [Fig.~\ref{fig:Pulse_shape_and_polarization}\,(h)].
    \textcolor{newtext}{The finite spatial extent of the acoustic pulse suppress one of the excited waves with larger $k$ [see Fig.~\ref{fig:Pulse_shape_and_polarization}\,(d)].
    This is a \textbf{Cherenkov-like} regime manifesting itself in $\mathbf{H}_{ext} \parallel \mathbf{k}$ geometry through the BVSW following the SAW wavepacket [Fig.~\ref{fig:SAW_MSW_Si_NiFe}\,(c,d)].}
    \textcolor{newtext}{Importantly, for a monochromatic acoustic wave the excited spin wave under the same conditions would remain confined within the acoustic beam and could not be radiated away, therefore, the Cherenkov‑like emission is enabled specifically by the short duration of the acoustic pulse.}
    \item The BW and SSW dispersion curves are tangent, with $v \approx v_{\text{c}}$ [Fig.~\ref{fig:Pulse_shape_and_polarization}\,(i)].
    \textcolor{newtext}{In this case, a propagating \textbf{magneto-elastic wavepacket} can form manifesting itself in the broadening of the BW wavepacket seen in the $\mathbf{H}_{ext} \perp \mathbf{k}$ geometry [Fig.~\ref{fig:SAW_MSW_Si_NiFe}\,(e,f)].
    Since in our system the BW pulse propagates predominantly in the substrate, its hybridization with spin waves in the thin Py layer represents a special case.
    For this reason, we clarify that by a magneto-elastic wavepacket we mean a coupled excitation whose dispersion near the crossing point of the acoustic and spin-wave branches differs from both uncoupled dispersions~\cite{gurevich2020magnetization}.}
    \item The L-SAW (R-SAW) dispersion has no shared points with BVSW (SSW) [$v_\text{n} < v_{\text{nc}}$, Fig.~\ref{fig:Pulse_shape_and_polarization}\,(k)]. 
    Hence, only \textbf{non-resonant} interaction is possible, enabling the detection of acoustic wavepackets with an enhanced signal-to-noise ratio via magneto-optical rotation [Fig.~\ref{fig:SAW_MSW_Si_NiFe}\,(c,e)].
    \textcolor{newtext}{
    Although this regime results in a low-amplitude forced magnetization dynamics, it can be exploited for distinguishing different acoustic modes, as we discuss in Sec.~\ref{sec:Selectivety}.
    }
\end{enumerate}

\textcolor{newtext}{
To conclude the theoretical consideration, we discuss the extent to which the two resonant regimes meet the conventional criteria for Cherenkov radiation and MEW.
One of the signatures of the Cherenkov mechanism is the radiation cone.
The formation of such cones in the magnon Cherenkov effect, for both excited waves with $k_1$ and $k_2$, was demonstrated in micromagnetic simulations~\cite{yan2013spin}.
In our system, the problem is quasi-one-dimensional owing to the small curvature of the acoustic wavefront, which prevents direct observation of the radiation cones.
For the MEW regime, the conventional anticrossing gap cannot be resolved for short acoustic pulses (duration of about one period). 
Instead, the formation of an MEW is identified by a change in the phase velocity of the acoustic packet upon resonant interaction with the magnetostatic wave [Fig.~\ref{fig:SAW_MSW_Si_NiFe}\,(f)].}

\textcolor{newtext}{\subsection{\label{sec:T-structure} \textbf{\textit{ In Situ}} Switching of Interaction Regimes in a T-shaped Waveguide}}

\textcolor{newtext}{Our analysis supported by the experimental data shows that, by tuning the difference between the group velocities of the acoustic and spin waves, one can achieve controlled switching between the regimes of magnetoacoustic interaction.
The dispersion of magnetostatic waves can be controlled by an external magnetic field.
Moreover, for magnonic device applications, such control must be performed locally.
This is possible in structures with a nonuniform magnetization direction, a classic example being the T-shaped waveguide~\cite{sadovnikov2015magnonic,martyshkin2024nonreciprocal,davies2015towards}.
Over a certain range of external field magnitudes and directions, the equilibrium magnetization rotates from the SSW to the BVSW geometry, enabling {\it in situ} observation of switching between interaction regimes during the propagation of an acoustic pulse.
Figure~\ref{fig:SAW_MSW_T-wg}\,(a) shows the sketch of the T-shaped waveguide used in the experiments (see Sec.~I of Suppl. Materials~\cite{Supplemental} for details).}

\begin{figure}
	\includegraphics[width=\linewidth]{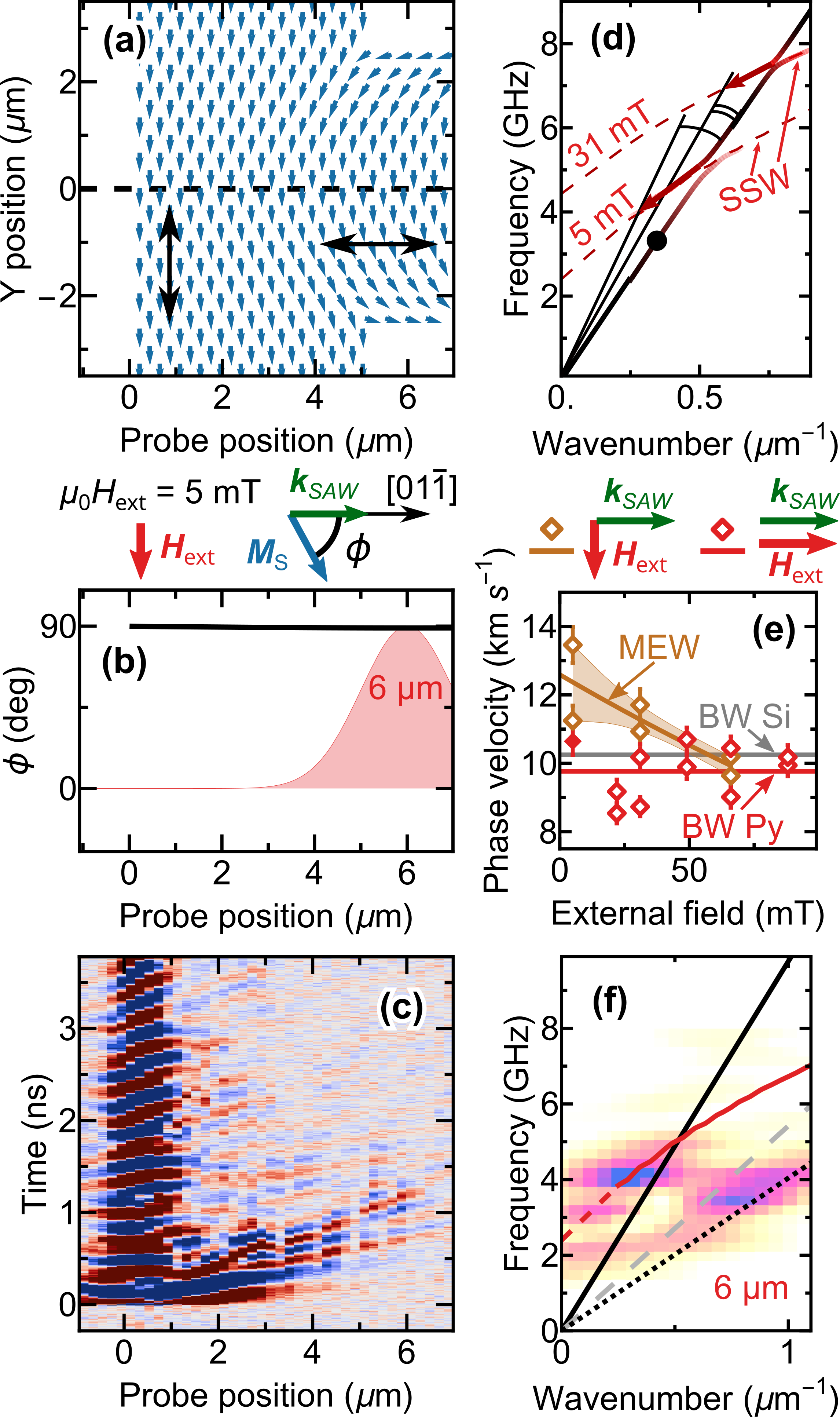}
	\caption{\label{fig:MEW}
		\textcolor{newtext}{{\bf Field-induced switching from MEW (II) to non-resonant regime (III).}
			(a) In-plane magnetization distribution in the T-waveguide at $\mu_0 H_{\text{ext}} = 5$\,mT ($\mathbf{H}_{ext}\perp\mathbf{k}$), obtained by micromagnetic simulation.
			The black line shows the scanning trajectory.
			(b) Magnetization deflection angle $\phi$ vs. distance from the waveguide edge along the scanning trajectory [black line in (a)].
			The red area marks the Gaussian spatial window (center at $x = 6$\,$\mu$m) used for local dispersion extraction.
			(c) 2D map of the pump-probe signal (polarization rotation) for the same geometry.
			(d) Modification of the phase velocity in the MEW regime.
			Red dashed lines: SSW dispersion at $\mu_0 H_{\text{ext}} = 5$ and $31$\,mT.
			The schematic anticrossing at the BW--SSW intersections illustrates how the phase velocity of the resulting MEW increases with decreasing field.
			(e) Measured phase velocity of the BW vs. external field for $\mathbf{H}_{\text{ext}} \parallel \mathbf{k}$ (red) and $\mathbf{H}_{\text{ext}} \perp \mathbf{k}$ (ocher).
			The shaded region marks the phase velocity spread within the pulse in the MEW regime. 
			The gray line shows the BW velocity on the bare Si substrate. 
			The closed symbol at $\mu_0 H_{\text{ext}} = 5$\,mT with $\mathbf{H}_{ext} \parallel \mathbf{k}$ is the phase velocity from the reflectivity measurement.
			(f) Local dispersion obtained from windowed 2D FFT of the data in (c) using the window shown in (b).
			Black (gray) lines correspond to the acoustic modes for which coupling to the magnetization is experimentally detected (absent) in the respective geometry; the red line shows the SSW dispersion.}
	}
\end{figure}

\textcolor{newtext}{First, we performed micromagnetic simulations to obtain the equilibrium magnetization distribution and the evolution of the magnetostatic-wave dispersion as a function of distance from the edge along the structure axis.
With an external field of $5$\,mT applied along the structure axis [Fig.~\ref{fig:SAW_MSW_T-wg}\,(a)], the magnetization rotates by $90\,^\circ$ along the waveguide [Fig.~\ref{fig:SAW_MSW_T-wg}\,(b,c)], because of the change in the local shape anisotropy [illustrated by the double headed arrows in Fig.~\ref{fig:SAW_MSW_T-wg}\,(b)].
The local magnetostatic-wave dispersion is extracted within the spatial Gaussian windows centered at $x_0=1$ and 6\,$\mu$m as marked in Fig.~\ref{fig:SAW_MSW_T-wg}\,(c) using the same approach as to the continuous film (Sec.~II in Suppl. Materials~\cite{Supplemental}).
The dispersion switches from an intermediate type between SSW and BVSW [red solid line in Fig.~\ref{fig:SAW_MSW_T-wg}\,(d)] to a pure BVSW (green line) type when the window is moved from $1$\,$\mu$m to $6$\,$\mu$m.
Thus, one expects that interaction regime between acoustic wavepackets and magnetostatic waves will be different depending on the distance along the waveguide structure.
In Fig.~\ref{fig:SAW_MSW_T-wg}\,(d) we also show the three acoustic wave dispersions and mark the positions of the spactral density maxima with black dots (see Sec.~III in Suppl. Materials~\cite{Supplemental}).}

\textcolor{newtext}{In the experiment, we successfully observe a change in the interaction regime during the propagation of the acoustic pulse [Fig.~\ref{fig:SAW_MSW_T-wg}\,(e)].
In the spatial range $0$--$3\,\mu$m, the dynamics is localized inside the acoustic pulse, which we identify as the non‑resonant regime.
For this range, we obtain the local experimental dispersion by applying a windowed two-dimensional Fast Fourier Transform to the spatio-temporal data of panel (e) using Gaussian windows with a FWHM of 2.4\,$\mu$m [shown as shaded areas in Fig.~\ref{fig:SAW_MSW_T-wg}\,(c)].
Local experimental and calculated dispersions are superimposed in Fig.~\ref{fig:SAW_MSW_T-wg}\,(f), showing that the non-resonant regime involves L-SAW pulse.
As can be seen, the crossing point of the L-SAW and spin-wave dispersions falls into the region $k \approx 0.8$\,$\mu$m$^{-1}$ where the acoustic packet amplitude drops significantly.
As a result, neither MEW nor Cherenkov regime of interaction is realized.}

\textcolor{newtext}{Starting at around $4\,\mu$m from the excitation area, long‑lived magnetization dynamics emerges after the acoustic packet, corresponding to the Cherenkov mechanism.
Within the area centered around $x_0=6$\,$\mu$m, the crossing point between L-SAW and BVSW dispersions is at $k \approx 0.45$\,$\mu$m$^{-1}$ [Fig.~\ref{fig:SAW_MSW_T-wg}\,(g)] corresponding to high spectral amplitude of the acoustic pulse, while the group‑velocity mismatch is large.
As a result of Cherenkov emission, the spectral density of the excited waves is localized near the dispersion crossing point.
When the field is increased, it overcomes the local shape anisotropy and the magnetization along the structure axis becomes nearly uniformly aligned with the external field.
In this case, the BVSW dispersion shifts to higher frequencies, the dispersions crossing point shift to the higher $k$ relative to the packet centre, and the non‑resonant regime dominates over the entire range of distances (Sec.~VII of Suppl. Material~\cite{Supplemental}).}

\textcolor{newtext}{When $\mathbf{H}_{\mathrm{ext}}$ is applied perpendicular to the structure axis, the magnetization along the axis remains uniform over the entire range of fields [see Fig.~\ref{fig:MEW}\,(a,b) and Sec.~VII of Suppl. Material~\cite{Supplemental}] and the magnetostatic wave disperion corresonds to that of SSW.
In this case, we observe the SAW wavepacket at all distances [Fig.~\ref{fig:MEW}\,(c)].
Near the excitation area, the ferromagnetic resonance contribution dominates, as is typical when the magnetization is aligned with an easy anisotropy axis~\cite{filatov2022spectrum} and we analyze interaction regimes only within the area centered at $x_0 = 6$\,$\mu$m.
The local dispersion reveals two wavepackets [Fig.~\ref{fig:MEW}\,(f)].
In the accessible range of $k$ at $\mu_0 H_{\mathrm{ext}} = 5$\,mT, only the SSW and BW dispersions intersect with $v \approx v_{\text{c}}$ [Fig.~\ref{fig:MEW}\,(d)].
One of the wavepackets corresponds to this crossing, suggesting that MEW regime is realized.
Another wavepacket corresponds to the non-resonant interaction with R-SAW.
For short MEW wavepackets, conventional detection of its signature -- an anticrossing splitting~\cite{carrara2024coherent,berk2019strongly,trzaskowska2022studies} is precluded by the strong spatial localization of the pulse and the resulting delocalization of its spectrum.
However, similar to the film case [Fig.~\ref{fig:SAW_MSW_Si_NiFe}\,(e,f)], we detect a significant increase in the phase velocity of the BW pulse (Sec.~VIII of Suppl. Material~\cite{Supplemental}).
Therefore, to substantiate the MEW nature of the wavepacket, we exploit the fact that the interaction modifies the SSW and BW dispersion curves, leading to a dependence of the wavepacket phase velocity on $H_{\mathrm{ext}}$ [Fig.~\ref{fig:MEW}\,(d) and Sec.~IX of Suppl. Material~\cite{Supplemental}].
Figure~\ref{fig:MEW}\,(e) shows the measured phase velocity of the BW packet for two directions of the external magnetic field -- along (red symbols) and perpendicular (ocher symbols) to the structure axis.
For the perpendicular orientation, where the magnetostatic waves exhibit the SSW dispersion and the condition $v \approx v_{\text{c}}$ is fulfilled, the phase velocity depends on the field magnitude.
As the field decreases, the phase velocity increases, and the spread among the velocities of different packet phases (shaded area) grows.
The latter indicates a deviation of the group velocity from the phase velocity.
At fields above approximately $50$\,mT, the crossing point shifts towards larger $k$, and the interaction switches to the non-resonant regime.
For the parallel orientation (BVSW geometry), no such field dependence is observed [red symbols in Fig.~\ref{fig:MEW}\,(e)].}

\begin{figure}
	\includegraphics[width=\linewidth]{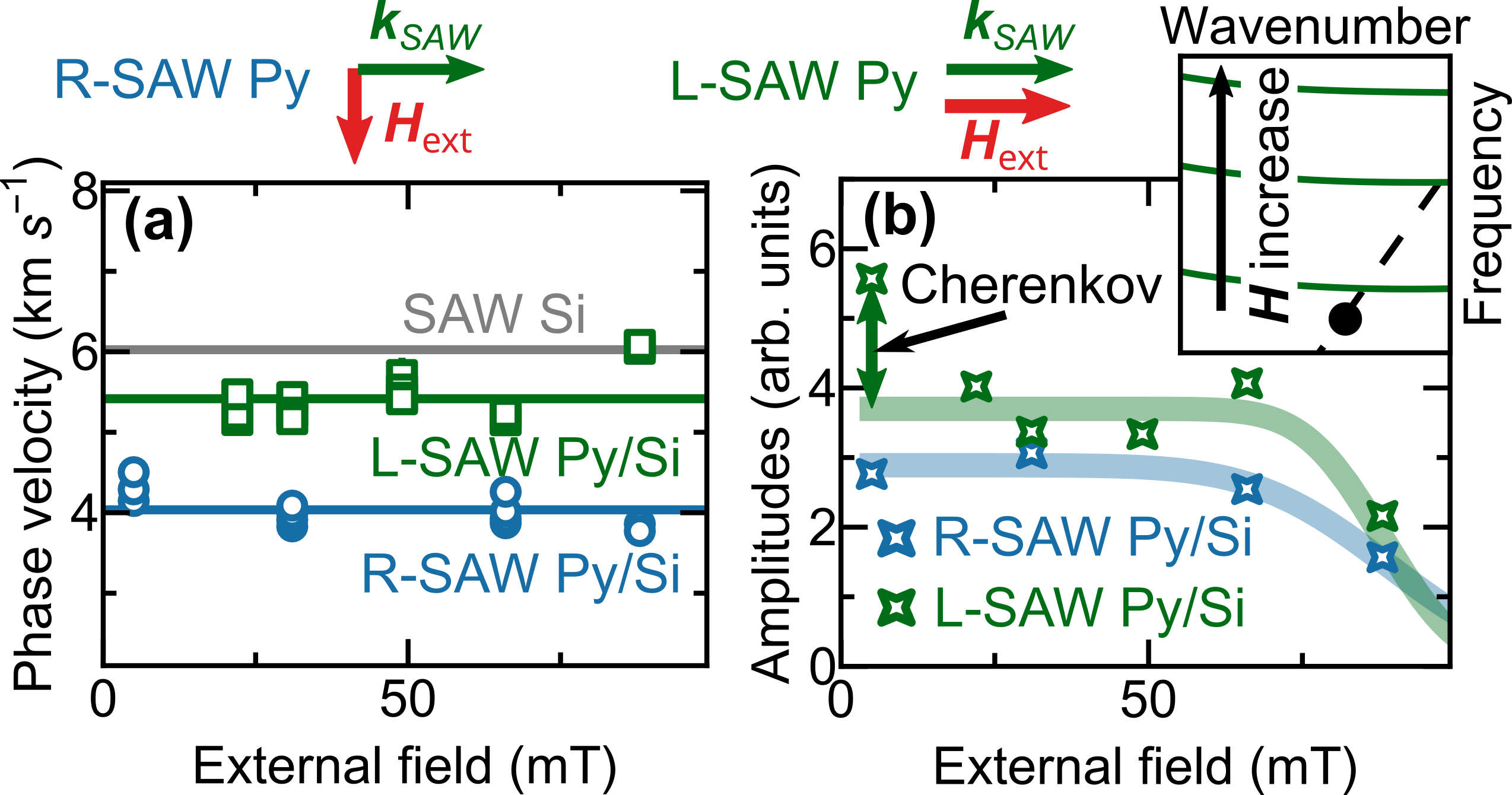}
	\caption{\label{fig:non-resonant}
		\textcolor{newtext}{{\bf Field-induced selectivity.}
			(a) Phase velocity of the SAW vs. external magnetic field for $\mathbf{H}_{\text{ext}} \parallel \mathbf{k}$ (green symbols) and $\mathbf{H}_{\text{ext}} \perp \mathbf{k}$ (blue symbols).
			The gray line indicates the SAW velocity on the bare Si substrate.
			(b) Integrated FFT amplitude within an area centered at $x = 6$\,$\mu$m vs. $H_{\text{ext}}$ for $\mathbf{H}_{\text{ext}} \parallel \mathbf{k}$ (green symbols) and $\mathbf{H}_{\text{ext}} \perp \mathbf{k}$ (blue symbols).
			At $H_{ext} > 88$\,mT, no signal could be detected in experiments.
			The double arrow marks the additional amplitude due to Cherenkov emission of BVSW.
			Lines are guides to the eye.
			The inset schematically shows the shift of the magnetostatic wave dispersion away from the acoustic packet center as the field increases.}
	}
\end{figure}

\textcolor{newtext}{\subsection{\label{sec:Selectivety} \textbf{Selective coupling between acoustic modes and magnetization dynamics}}}

\textcolor{newtext}{As shown above, the switching between regimes is governed by relative positions of the acoustic and magnetostatic wave dispersions, group velocity mismatch at the crossing point, and the spectral density of the acoustic pulse at the wavevectors correspondign to the crossing point.
This naturally creates the selectivity of the two resonant regimes of interactions.}

\textcolor{newtext}{
The selectivity of the non-resonant regime is expected to be related to the detuning of the dispersions.
In experiment we observe non-resonant coupling of BVSW to L-SAW and SSW - to R-SAW.
This selectivity is deduced from the SAW phase velocities obtained in the waveguide within area around $x_0 = 6$\,$\mu$m when field is applied along and perpendicular to the waveguide axis as shown in Fig.~\ref{fig:non-resonant}\,(a).
In the non-resonant regime, the measured phase velocities coincide with the group velocities of the acoustic wavepackets.
Clearly, that in the BVSW and SSW geometries, SAWs with different velocities are detected, corresponding to L-SAW and R-SAW respectively.
However, as illustrated in Fig.~\ref{fig:Pulse_shape_and_polarization}\,(k), this selectivity cannot be related to the dispersions detuning.
We attribute the emergence of this selectivity to the features of the driving force inserted on magnetization by short laser-induced acoustic pulses as detailed in Sec.~VI of Suppl. Material~\cite{Supplemental}.
The fact that the SAW wavepackes are detected through their non-resonant coupling to the corresponding magnetostatic wave is confirmed by the field dependence of the signal amplitude [Fig.~\ref{fig:non-resonant}\,(b)].
The amplitude was determined as the integral of the spectral density of the signal at area around $x_0 = 6\,\mu$m, where only the signal from propagating wavepackets or waves is present.
In the geometry with the external field along the T-structure axis, a peak is observed at $\mu_0 H_{\mathrm{ext}} = 5$\,mT, corresponding to the additional contribution from the Cherenkov emission [Fig.~\ref{fig:SAW_MSW_T-wg}\,(e)].
Under any other field magnitude and direction, only non-resonant interaction is expected and the signal amplitude is indeed lower.
As the field increases, the magnetostatic-wave dispersions in both geometries shift toward higher frequencies and the efficiency of non-resonant interaction regime diminishes [see the schematic in the inset of Fig.~\ref{fig:non-resonant}\,(b)].
Indeed, further increase to $66$\,mT results in the amplitude drop, and for fields above $88$\,mT the signal from the propagating wavepackets is no longer detected.
This field dependence of the amplitude confirms that L-SAW and R-SAW are detected through their non-resonant coupling to magnetostatic waves.
The observed effect can be used for selective detection of the L- or R‑SAW.}

\textcolor{newtext}{\section{\label{sec:concl} Conclusions and Outlook}}

\textcolor{newtext}{In summary, we experimentally realized three regimes of picosecond magnetoacoustic interaction -- Cherenkov-like spin-wave emission, magnetoelastic wavepacket formation, and non-resonant forced magnetization oscillations -- within a single T-shaped waveguide using picosecond acoustic wavepackets.
Using a theoretical model, we established the criterion for the transitions between these regimes, namely the detuning between the group velocities of the acoustic wavepacket and magnetostatic waves.
Crucially, we demonstrated that switching between regimes can be achieved locally during the propagation of a single acoustic pulse by exploiting a nonuniform equilibrium magnetization distribution.
As compared to conventional magnetoacoustics, using laser-driven short acoustic wavepackets enable observation of Cherenkov emission.
Formation of MEW is facilitated owing to localization of the magnetization dynamics around the acoustic pulse.
Finally, selectivity of the non-resonant regime may also result from the short duration of the acoustic wavepacket.
Our results not only clarify the conditions governing different picosecond magnetoacoustic regimes but also pave the way toward reconfigurable magnonic devices with enhanced energy efficiency.}

\textcolor{newtext}{The demonstrated regimes of picosecond magnetoacoustic interaction open a pathway for implementing magnonic logic and improving its energy efficiency.
The low attenuation and long propagation length of SAWs offers a solution for the problem of spin-wave damping during transmission between devices.
A composite wavepacket (a propagating SAW pulse accompanied by a Cherenkov-like magnetization dynamics) can serve as a phase-coherent computational object.
Its long spin-wave tail enables reliable interference, which is suitable for binary logic, where constructive and destructive interference encodes logical states~\cite{mahmoud2020introduction}.
Controlled phase shifts, essential for operations, can be introduced by locally switching the interaction to the MEW regime.
Furthermore, the steady-state spin-wave amplitude, governed by damping, provides intrinsic signal normalization.
This allows the composite wavepacket to be utilized as a long-range signal carrier, mitigating the severe attenuation problem of pure spin waves and addressing the energy-efficiency challenge highlighted for magnonic computing.}

\begin{acknowledgments}
The work of P.I.G. on pump-probe experiments, theoretical analysis, and micromagnetic simulations was supported by the grant of the RSF No. 24-72-00136, https://rscf.ru/project/24-72-00136/.
The sample technology development and fabrication were done by N.S.G. and M.V.S. in the frames of State Contract No. FFUF-2024-0021 using the facilities of Center “Physics and Technology of Micro and Nanostructures” at IPM RAS.
We thank A.~Yu.~Klokov for providing us with scripts for SAW dispersion calculation.
\end{acknowledgments}

\normalem% Remove underline for books in bibliography
\bibliography{apssamp.bib}% Produces the bibliography via BibTeX.

\end{document}